\documentclass[notitlepage, twocolumn, letterpaper, prx, longbibliography, 10pt]{revtex4-2}

\usepackage{graphicx}% Include figure files
\usepackage{dcolumn}% Align table columns on decimal point
\usepackage{bm}% bold math
\usepackage[utf8]{inputenc}
\usepackage[T1]{fontenc}
\usepackage{mathptmx}
\usepackage{etoolbox}
\usepackage{amsmath}
\usepackage{amssymb}
\usepackage{hyperref}
\usepackage{subcaption}
\usepackage[percent]{overpic}
\usepackage[normalem]{ulem}
\usepackage[switch,columnwise]{lineno}
\usepackage[object=vectorian]{pgfornament} % fancy hrules
\usepackage[outdir=./]{epstopdf}
\usepackage{tikz-cd}

\newcommand{\dsm}[1]{{\textcolor{teal} {({\tiny DSM:} #1)}}}

\newcommand\numberthis{\addtocounter{equation}{1}\tag{\theequation}}

\newcommand{\dd}[1]{\mathrm{d}#1}

\newcommand{\deriv}[2]{\frac{\dd{#1}}{\dd{#2}}}

\begin{document}

% \preprint{AIP/123-QED}

\title[Mexican hat clustering]{A Mexican hat dance: clustering in Ricker-potential particle systems}
% Force line breaks with \\
\author{D. Sabin-Miller}
\email{dasami@umich.edu}
\affiliation{University of Michigan, Center for the Study of Complex Systems}
 %\altaffiliation[Also at ]{Physics Department, XYZ University.}%Lines break automatically or can be forced with \\
\author{D. Abrams}%
 %\email{dmabrams@northwestern.edu.}
\affiliation{ 
Northwestern University, Engineering Sciences and Applied Mathematics%\\This line break forced with \textbackslash\textbackslash
}%

% \author{C. Author}
%  \homepage{http://www.Second.institution.edu/~Charlie.Author.}
% \affiliation{%
% Second institution and/or address%\\This line break forced% with \\
% }%
%\graphicspath{{images/}}

\date{\today}% It is always \today, today,
             %  but any date may be explicitly specified

\begin{abstract}
The dynamics and spontaneous organization of coupled particles is a classic problem in modeling and applied mathematics. Here we examine the behavior of particles coupled by the Ricker potential, exhibiting finite local repulsion transitioning to distal attraction, leading to an energy-minimizing ``preferred distance''. When compressed by a background potential well of varying severity, these particles exhibit intricate self-organization into ``stacks" with varying sizes and positions. We examine bifurcations of these high-dimensional arrangements, yielding tantalizing glimpses into a rich dynamical zoo of behavior.
%The behavior and spontaneous organization of particles exhibiting a finite local repulsive force and an energy-minimizing ``preferred distance'' is useful. \dsm{why?} In physics, the Morse potential is of this form \dsm{Lennard-Jones (difference between twelfth and sixth inverse powers) has infinite repulsion. Morse is the one we want, as long as we symmetrize it, but that's not differentiable at 0 and the sharp peak ruins clumping}. Here we examine a simpler analogous system of particles with Ricker potentials, and find counterintuitive clustering behavior when these particles are confined in a quadratic potential well.
%Connect to Lennard-Jones and Morse potentials, mention Ricker potential as more tractable analogue, and then counterintutive clustering of particles under pressure.
\end{abstract}

\maketitle

\section{Introduction} \label{sec:intro}

%A classic problem in physics and applied mathematics concerns the behavior of a system of interacting particles constrained by a confining potential \dsm{cite classic oscillators, particles-in-a-box, Bose-Einstein condensates, potential-coupling work(?)}. Here we investigate such a system where the particles exhibit short-range repulsion and long-range attraction, a common type of interaction often referred to as a ``Mexican hat'' potential \dma{cite} (see Fig.~\ref{fig:hat}).  Particle interaction of this type occurs in nature, e.g., via the Lennard-Jones and Morse potentials, and can be modeled qualitatively via the Ricker wavelet \dma{cite each}.  

%\dsm{now repeats some of the abstract}
Systems of coupled particles are a classic problem in physics and applied mathematics (e.g.,~\cite{slater1951simplification, kuramoto1975self, csahok1997transport, schwerdtfeger2006extension} and many others). %\dsm{many-body ?Lennard-Jones? 'soft-core' potential problems in physics, Hartree-Fock}
%) %\dsm{cite classic oscillators, potential-coupling work}
Here we investigate particles interacting via short-distance repulsion and long-distance attraction, often referred to as a ``Mexican hat'' potential %\dsm{cite}
(see Fig.~\ref{fig:hat}). This qualitative scenario is seen in intermolecular forces and can be modeled via, e.g., the Lennard-Jones and Morse potentials. 

Here we explore the behavior as one-dimensional populations of such particles are ``squeezed'' together, similar in concept to ``particle in a box'' considerations from quantum physics (e.g.,~\cite{cohen2019quantum}).

% The primary domain of examination here is the infinite real line equipped with a quadratic potential well, but in the Appendix we also present preliminary findings for an ``oscillator'' interpretation with the particles' positions on a periodic domain representing phase (see section \ref{sec:Oscillators}).

% For simplicity, however, we utilize the Ricker wavelet (as explored in, e.g.,~\cite{Hosken_1988_Ricker,Wang_2014_Ricker}) as our potential function. 

% Work has been done on the birth of multimodality in similar systems \cite{johnson2019coupled}, but 

%\dma{Now rephrase this paragraph a bit:} 
Local repulsion and distal attraction may call to mind the Lennard-Jones and Morse potentials from physics (see section \ref{sec:LJ_and_Morse}). These models for intermolecular potential energy have features rendering them distinct from our Ricker wavelet: infinite repulsion in the case of Lennard-Jones, and a ``sharp'' non-differentiable peak at the origin for Morse. However, these models may not be applicable in situations where coexistence at the same position is allowed, due to their nonphysical implications at $x=0$. However, we believe a ``smoothing'' of the Morse potential's central peak (such as by integration against a ``blurring'' kernel function) would cause qualitatively similar results to what we observe in our Ricker system, and it is possible other ``soft-core'' potential systems (e.g., \cite{Maessen1989soft,rijken1999soft,Yatsenko2004soft}) could find our results applicable. The smooth and coexistence-friendly dynamic embodied by our Ricker wavelet may also apply to neuronal phase models under proper conditions (e.g., \cite{ermentrout1986parabolic,ermentrout1996type,hoppensteadt1997weakly}).% with both inhibitory and excitatory interactions   \dma{
%I'd mention it and say something like ``...could apply to neuronal models like the integrate-and-fire model [cite] if both inhibitory and excitatory interactions are included.'' 
%See https://neuronaldynamics.epfl.ch/online/Ch5.S4.html for some refs. 140, 141, 228}.

% \section{Motivation} \label{sec:intro}

% \dma{need to rephrase this}Prior work has been done on the birth of multimodality in similar systems \dsm{cite Joe}, but we wanted to explore the behavior as populations of such particles were ``squeezed'' together. Here we examine the effect of a quadratic potential well on the infinite real line, though in the appendix we also consider ``oscillator'' interpretations by embedding the particles on a periodic domain.

\section{The Modified Ricker Potential}
\begin{figure}[t!] 
\centering
	\includegraphics[width=.9\columnwidth]{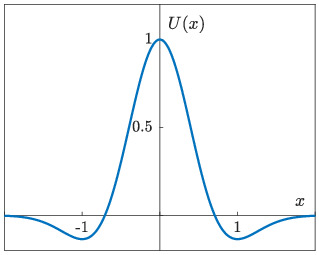}
	\caption{\textbf{Ricker wavelet.} Interaction potential $U(x)$ for a particle described by the Ricker wavelet potential (Eq.~\eqref{eq:hat}) with parameters $s=1$ and $k=2$.}
	\label{fig:hat}
\end{figure}

We use a modified form of the Ricker wavelet as the potential function carried or ``worn'' by each particle,
\begin{equation}
    U(x) = \left[ 1-\frac{k}{k-1} \left(\frac{x}{s} \right)^2\right]e^{-k\left( \frac{x}{s}\right)^2},
    \label{eq:hat}
\end{equation}
which is pictured in Fig.~\ref{fig:hat}.

This function has the following properties: 
\begin{enumerate}
    \item Central peak at $(0,1)$
    \item Symmetric troughs (i.e., local minima) at $x = \pm s$
    \item Trough depth controlled by $k \in [1,\infty)$: as $k\rightarrow 1^+$, trough depth $\rightarrow -\infty$, and as $k \rightarrow \infty$, trough depth $\rightarrow 0^-$ 
\end{enumerate}
Due to the central hump and symmetric troughs, this potential provides short-range, finite repulsion coupled with long-range attraction to a ``preferred separation'' $s$. 
%This qualitative shape qualifies it as a ``Mexican Hat Potential".

The potential at position $x$ due to a particle at position $x_i$ is
\begin{equation}
    U(x|x_i) = \left[ 1-\frac{k}{k-1} \left(\frac{x-x_i}{s}\right)^2\right]e^{ -k\left(\frac{x-x_i}{s}\right)^2}. \label{eq:Ricker_potential_xi}
\end{equation}

%%%%%%%% outline thoughts %%%%%%%%%%
%Coupled oscillators - frozen disorder, or quasi-stability? Observations about global potential high-frequency damping
%Morse Potential? nope, qualitatively different stability with sharp origin peak

%possible orders: introduce system then
% 1) intriguing self-organizing results
% 2) small-n analysis and figures
% 3) large-n analysis

% 1) small-n analysis
% 2) large-n intriguing results
% 3) large-n analysis

% 4) Oscillator interpretation, high-frequency-damping to quasi-stable states?

\begin{figure}[t!] 
\centering
	\includegraphics[width=.9\columnwidth]{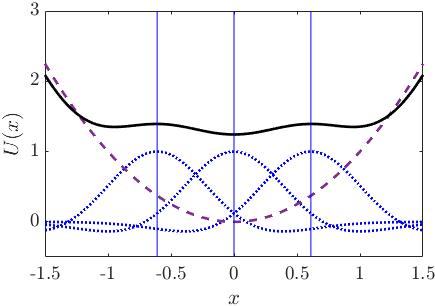}
	\caption{\textbf{Three particles at equilibrium.}  Example of three particles at an equilibrium. The vertical solid blue lines show particle positions, the dotted blue curves show the particles' potentials, and the dashed maroon parabola is the background potential well. The solid black curve is the total potential, and we can see the derivative is zero at each particle's position, indicating this arrangement is at equilibrium. This arrangement is stable: since a particle does not influences itself, each particle effectively ``sees'' the global potential minus its own contribution, which makes each particle's position in this arrangement a ``trough'' from its own point of view. } 
	\label{fig:ex_potentials_in_well}
\end{figure}

We suppose that $n$ particles, indexed $1$ through $n$, have one-dimensional positions $x_i$ and influence each other through their modified Ricker potential via the first order dynamical system
\begin{equation*}
    \deriv{x_i}{t} %&= -\deriv{V}{x}\biggr\rvert_{x_i}\\
                   = -\left.\deriv{U_{tot}}{x}\right|_{x_i} = -\left(\deriv{U_0(x)}{x} + \sum_{j=1}^n \deriv{U(x|x_j)}{x} \right) \Biggr\rvert_{x_i} ,
\end{equation*}
where confinement is imposed by the global potential function $U_0$, which we assume for simplicity to have a quadratic shape
$$
U_0(x) = \frac{x^2}{w^2}
$$
with width-control parameter $w$. Figure \ref{fig:ex_potentials_in_well} shows an example arrangement of particles in such a system.  

We note for later reference the first and second derivatives of our Ricker potential:
\begin{equation}
   \deriv{U}{x} = -\frac{2k^2}{(k-1)s^2}x\left( 1 - \frac{x^2}{s^2} \right) e^{-\frac{kx^2}{s^2}},
   \label{eq:dRicker}
\end{equation}
\begin{equation}
    \frac{\textrm{d}^2U}{\textrm{d}x^2} = -\frac{2k^2}{(k-1)s^2}\left[ 1 - \frac{3+2k}{s^2}x^2  + \frac{2k}{s^4}x^4  \right] e^{-\frac{kx^2}{s^2}}\;.
    \label{eq:d2Ricker} 
\end{equation}
In particular we note that due to the zero derivative of the Ricker potential at the origin, there is no need to complicate notation by explicitly excluding particles from influencing themselves; particles have no self-interaction regardless.

Without loss of generality, we henceforth restrict our analysis to the case $s=1$, since an appropriate rescaling of space and time ($\Tilde{x} = x/s$, $\Tilde{w} = w/s$, $\Tilde{t} = t/s$) removes that parameter from the governing equations.  The parameter $k$ does qualitatively affect system behavior (which we briefly explore in Section \dsm{fix}S.2 of the Supplementary Materials (SM)), though our analysis is primarily concerned with the confinement parameter $w$.  Unless otherwise noted, our numerical examples use the default parameter value $k=2$.

%Unless otherwise noted, in our numerical examples we use default parameter values $s=1, k=2$. We note that $s$ may be scaled out of the problem entirely with appropriate scaling of space (via $\Tilde{x} = x/s$ and $\Tilde{w} = w/s$) and time (as $\Tilde{t} = t/s$), but the parameter $k$ does qualitatively affect behavior (as we briefly explore in Section \dsm{fix}S.2 of the Supplementary Materials (SM)).

% \section{1D Quadratic Potential Well}
% We suppose the particles are contained in the background potential well
% $$
% U_0(x) = \frac{x^2}{w^2}
% $$
% with width-control parameter $w$.

\subsection{Intriguing Collective Behavior}
Free of confinement (i.e., with $w\to \infty$) and in the absence of any degenerate initial positions, particles will settle into a state of uniform spacing, with each particle residing at the preferred distance $s$ from its neighbors. However, when confined, they exhibit highly nonuniform and rich behavior.
%\dma{I'd start with something like ``One might expect particles with locally repulsive interaction to space out uniformly at equilibrium, and that is exactly what we observe as the confinement is relaxed ($w \to \infty$).  However, when $w$ shrinks and confinement effects become important, surprisingly rich behavior is observed.}

\begin{figure}[t!] 
\centering
	\includegraphics[width=.9\columnwidth]{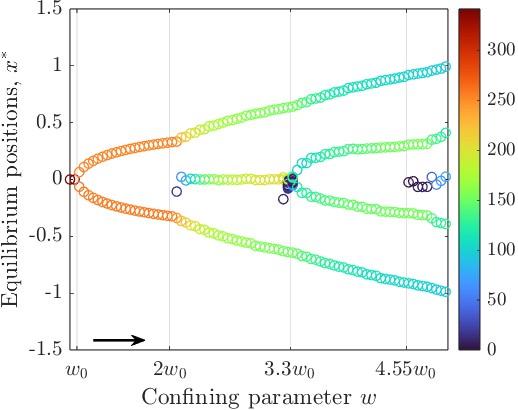}
	\caption{\textbf{Equilibrium diagram.} Apparently stable equilibrium positions for $512$ particles confined in a quadratic potential well. Color indicates particle abundance. The horizontal $w$-axis tick marks are placed at approximate bifurcation points, and persist on other diagrams of this type for comparison's sake. The horizontal scale is set by $w_0= \frac{s}{k}\sqrt{\frac{k-1}{n}}$, which is the critical $w$ value where the fully-stacked origin state becomes unstable (see section \ref{sec:general-n_stability}).
    %---a value which depends on the Ricker parameters $s$ and $k$, and the number of particles $n$. 
    %Arrow indicates that that equilibria were continued with gradually increasing parameter $w$ (along with minute perturbations to seek stable states only). Ricker-potential constants for this run (and all other diagrams in this section) were $s=1, k=2$. 
    See section \ref{sec:methods} for additional simulation details. 
    %\dma{maybe start figure at $w=0$?}
    }
	\label{fig:512_particles}
\end{figure}

When confinement is present, we observe spontaneous organization of the particles depending on the choice of confinement parameter $w$ (note that confinement is stronger as $w$ decreases toward zero). Particles form ``stacks''---states where multiple particles occupy the same spatial position---since their repulsion weakens as they get nearer to each other, but the number of particles in each stack can vary, and indeed the stacks exchange particles as $w$ changes. We demonstrate this spontaneous organization in Fig.~\ref{fig:512_particles}, which shows the results of numerical investigation of the system's stable equilibria.

\begin{figure*}[t!] 
\centering
	\includegraphics[width=.45\textwidth]{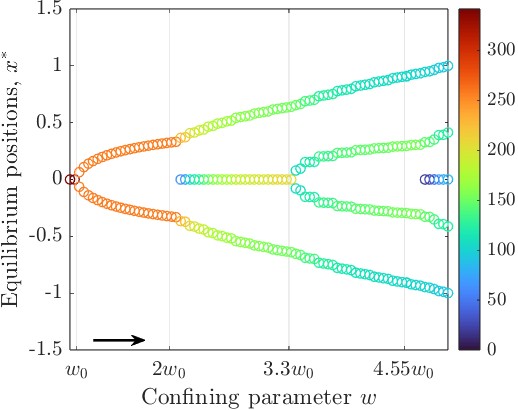}
    \includegraphics[width=.5\textwidth]{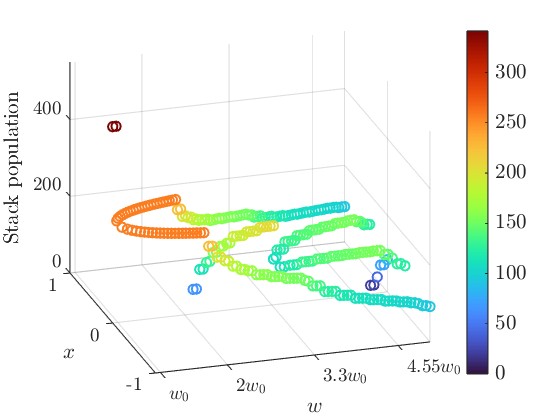}
	\caption{\textbf{Equilibrium diagram, $n=512$, symmetry enforced.} \textbf{Left:}A symmetry-enforced version of Fig.~\ref{fig:512_particles}. Besides being ``cleaner,'' however, we also notice the apparent bifurcation points change slightly due to the minimum of two particles in a stack. \textbf{Right:} A 3D view of the data in the left panel, with stack-size information encoded in the vertical axis as well as color. This emphasizes the continuous shift in population fractionation and the structure of the major bifurcations.} 
	\label{fig:256x2_F_w_3D}
\end{figure*}

Note that there is some slight asymmetry in Fig.~\ref{fig:512_particles} due to high dimensional multistability with various stack sizes; if we enforce symmetry, we see a picture of ``core'' behavior as shown in Fig.~\ref{fig:256x2_F_w_3D}.

\begin{figure*}[t!] 
\centering
	\includegraphics[width=.34\textwidth]{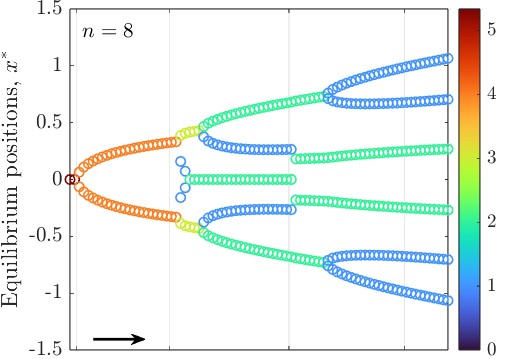}
	\includegraphics[width=.3\textwidth]{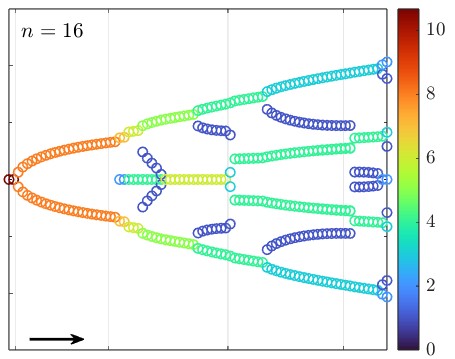}
	\includegraphics[width=.3\textwidth]{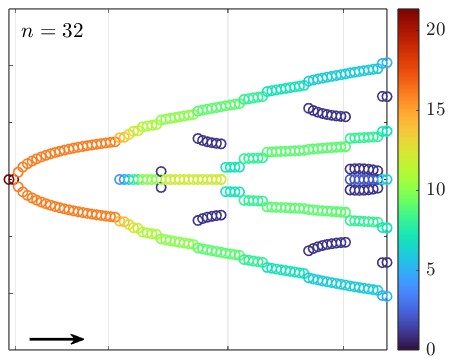} \\
	\noindent
	\centering
	\includegraphics[width=.34\textwidth]{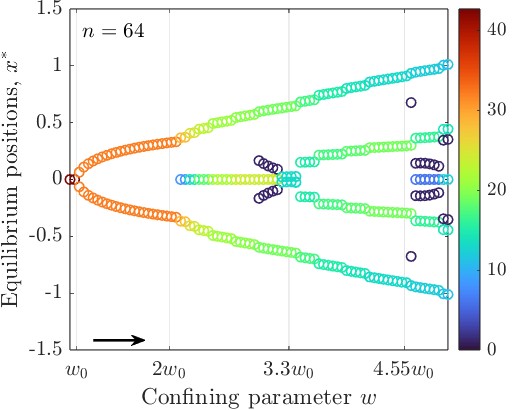}
	\includegraphics[width=.3\textwidth]{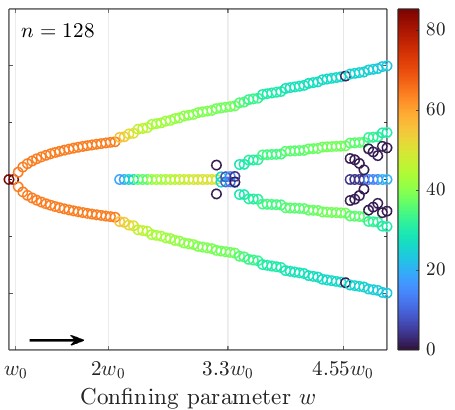}
	\includegraphics[width=.3\textwidth]{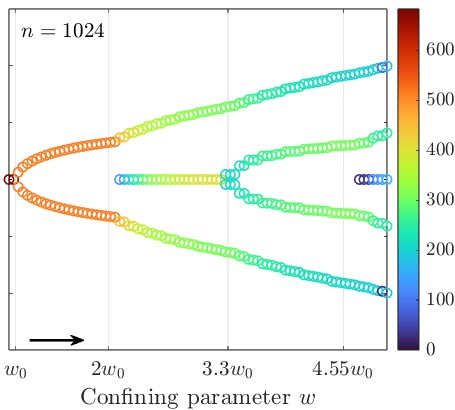}
	\caption{\textbf{Symmetry-enforced system, $n$ convergence.}  Equilibrium diagrams for increasing population sizes $n$. The diagrams appear to converge in a visual sense to the same few ``major'' bifurcations, which occur at nearly the same multiples of the critical parameter value $w_0$ (given in Eq.~\eqref{eq:w0_general}).  This motivates us to understand and characterize this large-$n$ generic behavior.} 
	\label{fig:colorscatter_convergence}
\end{figure*}

Also, note that these figures actually display $2$-dimensional projections of an $n$-dimensional bifurcation diagram, with each particle's position occupying one dimension. However, the interchangeability symmetry of the particles allows the display of the whole population to serve as analogue for any single particle's possible positions. 

The stack-size information (encoded by color) can be further emphasized with a third dimension, as shown in the right panel of Fig.~\ref{fig:256x2_F_w_3D}.
% \begin{figure}[ht!] 
% \centering
% 	\includegraphics[width=.8\columnwidth]{256x2_3d_F.eps}
% 	\caption{\textbf{Symmetry-enforced large system, 3D view.} A 3D view of the data in Fig.~\ref{fig:256x2_F}, with stack-size information encoded in the vertical axis as well as color. This emphasizes the continuous shift in population fractionation and the structure of the major bifurcations.  } 
% 	\label{fig:256x2_F_3D}
% \end{figure}

Thus we see even more clearly the pattern of stable behavior. For very small $w$ (strong confinement, i.e.,~a narrow parabolic well) all the particles stack up at the origin, but the population splits apart into two symmetric stacks\footnote{There is some multistability with regards to slight asymmetry of these two stacks; for example, sometimes the stacks are 63 and 65 particles, and the system is stable, although in that case their positions are not perfectly symmetric---the larger stack settles slightly closer to the origin.} when $w$ passes a critical threshold we label $w_0$---we derive a formula for this value (Eq.~\eqref{eq:w0_general}) in the ``Large $n$ analysis'' section. As $w$ continues to grow, the stacks drift apart and the origin becomes stable again, and we see particles ``fall'' inwards to settle there. Initially only a few particles stably rest there, but as the stacks continue to separate the central stack grows, until it becomes large enough to split into two in a manner that appears similar to its initial bifurcation at $w=w_0$. 

There are many other, more complex equilibria possible, but for large $n$ the equilibrium diagram appears to become increasingly well characterized by %become dominated by %well characterized primarily by converge to 
the aforementioned behavior, as shown in Fig.~\ref{fig:colorscatter_convergence}. All the ``major'' bifurcations (birth of central stacks, splitting of central stacks) appear to happen at the same multiples of the critical parameter value $w_0$, thus with proper scaling of the $w$ axis (to match $w_0$) the diagrams appear increasingly similar to one another.  

%\dsm{maybe more exotic behavior possible, like outer branches branching...?}

We will start our exploration with smaller, more tractable examples and then progress from there to the more general, large-$n$ cases.

\section{Small-$n$ particular case analysis}
\subsection{Two-particle case}
We can solve for the equilibrium condition on the right particle by assuming the particles are at $\pm x^*$ (plugging in $2x^*$ as the distance in equation \eqref{eq:dRicker} and adding the background potential at position $x^*$):
\begin{align}
    \deriv{U_\textrm{tot}}{x}\biggr \rvert_{x=x^*} &= \frac{2x^*}{w^2} - \frac{4k^2x^*}{(k-1)s^2}\left( 1 - 4\frac{x^{*2}}{s^2} \right) e^{-\frac{4kx^{*2}}{s^2}}\;. \nonumber
\end{align}
At equilibrium this slope is zero, thus $x^*=0$ (corresponding to both particles stacked at the origin) or    
          %0 &= 2x^* \left[ \frac{1}{w^2}e^{\frac{4kx^{*2}}{s^2}} - \frac{2k^2}{(k-1)s^2}\left( 1 - 4\frac{x^{*2}}{s^2} \right)\right] e^{-\frac{4kx^{*2}}{s^2}} \nonumber \\
%          \implies x^* &= 0 \qquad \textrm{or} \nonumber\\
\begin{equation}
    0 = \frac{1}{w^2}e^{\frac{4kx^{*2}}{s^2}} - \frac{2k^2}{(k-1)s^2}\left( 1 - 4\frac{x^{*2}}{s^2} \right) \label{eq:2_particle_equil}    
\end{equation}

\begin{figure}[t!] 
\centering
	\includegraphics[width=.8 \columnwidth]{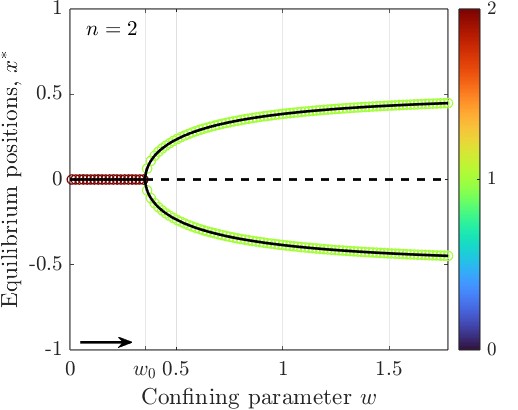}
	\caption{\textbf{Two-particle bifurcation diagram, with exact solutions.} Equilibria for the two-particle system, with exact analytical solutions $x^*=0$ (dashed when unstable) and Eq.~\eqref{eq:2_particle_equil} overlaid in black. The critical parameter value $w_0$ ($=\frac{1}{\sqrt{8}}$ in this case) is marked as well. } 
	\label{fig:2_particles}
\end{figure}

%\end{align}
So the bifurcation diagram (in $w, k,$ or $s$) is given by the implicit equation \eqref{eq:2_particle_equil}. This corresponds to the exact solution
\begin{equation}
    x^*(w) = \frac{1}{2} \sqrt{1 - \frac{1}{k}W\left(\frac{(k-1)e^k}{2 k w^2}\right)}
\end{equation}
\begin{equation}
    x^*(w) = \frac{s}{2} \sqrt{1 - \frac{1}{k}W\left(\frac{1}{2}\frac{k-1}{k}\frac{s^2}{w^2} e^k\right)}
\end{equation}
%\dma{Double check: What was $n$? Also decide if we want to eliminate $s$ from all equations or not.}
where $W$ is the Lambert W function, defined as the solution to 
\[
    W(z) e^{W(z)} = z\;.
\]
Fig.~\ref{fig:2_particles} shows this solution overlaid on the empirical equilibrium diagram for two particles.

An approximation for $|x| \ll 0$ yields 
%the bifurcation point $w_{0,2}$ (the initial bifurcation $w_0$ with $n=2$) and the shapes of the branches of solution in its neighborhood:
%
%For a more explicit, albeit approximate solution, we may expand \eqref{eq:2_particle_equil} for small $x^*$ (effectively assuming near-bifurcation $w$):
%
% \begin{align*}
%     x &= \epsilon \\
%     0 &= \frac{1}{w^2}\left(1+\frac{4k\epsilon^2}{s^2} + O(\epsilon^4) \right) \\
%     & \qquad \qquad - \frac{2k^2}{(k-1)s^2}\left( 1 - 4\frac{\epsilon^2}{s^2} \right) \\
%     %0 &=\left(1-\frac{2w^2k^2}{s^2(k-1)}\right)+\left(\frac{4k}{s^2} +\frac{2w^2k^2}{s^2(k-1)} \right)\epsilon^2+ \frac{8k^2}{s^4}\epsilon^4 + O(\epsilon^6)\\
%     \frac{2k^2w^2}{(k-1)s^2}\left( 1 - 4\frac{\epsilon^2}{s^2} \right) &= 1+\frac{4k\epsilon^2}{s^2} + O(\epsilon^4)  \\
%     \frac{2k^2w^2}{(k-1)s^2} &= \frac{1+\frac{4k\epsilon^2}{s^2} +  O(\epsilon^4)}{\left( 1 - 4\frac{\epsilon^2}{s^2} \right)}\\
%     &= \left(1+\frac{4k\epsilon^2}{s^2} + O(\epsilon^4) \right) \left(1 + 4\frac{\epsilon^2}{s^2} + O(\epsilon^4)\right) \\
%     &= 1 + \frac{4+4k}{s^2}\epsilon^2 +  O(\epsilon^4) \\
%     w^2 &= \frac{(k-1)s^2}{2k^2}  \left( 1 + 4\frac{k+1}{s^2}\epsilon^2  + O(\epsilon^4) \right)\\
%     w(x) &\sim \frac{s}{k}\sqrt{\frac{k-1}{2}}\left(1 + 2\frac{k+1}{s^2}x^2 +  O(x^4) \right) \numberthis
% \end{align*}
%Then by inversion, defining the constant $w_{0,2}$, we get the leading-order $x(w)$ behavior near this bifurcation:
\begin{equation}
    x^*(w) \approx \frac{s}{\sqrt{2w_{0,2}(k+1)}} \sqrt{w-w_{0,2}} \;,
    \label{eq:xapprox}    
\end{equation}
where we have defined $w_{0,2} = \frac{s}{k}\sqrt{\frac{k-1}{2}}$ as the lowest critical value of $w$ with $n=2$.  We derive the general-$n$ version (which we simply refer to as $w_0$ going forward) in the ``Large-$n$ analysis'' section (Eq.~\eqref{eq:w0_general}).

% \begin{align} 
%     w_{0,2} &:= \frac{s}{k}\sqrt{\frac{k-1}{2}} \label{eq:w0_n=2} \\
%     x^*(w)&\phantom{:}\sim \frac{s}{\sqrt{2w_{0,2}(k+1)}} \sqrt{w-w_{0,2}} \;.\label{eq:xapprox}
% \end{align}
% This matches the expansion of the exact Lambert-W-based solution near $w=w_{0,2}$. Eq.~\eqref{eq:w0_n=2} gives $w_0$ for the $n=2$ case; we derive the general-$n$ version in the ``Large-$n$ analysis'' section (Eq.~\eqref{eq:w0_general}).
% Then by inversion, using the constant $w_0$:
% \begin{align} \label{eq:xapprox}
%     w_0 &= \frac{s}{k}\sqrt{\frac{k-1}{2}} \nonumber \\
%     x(w) &\sim \frac{s}{\sqrt{2w_0(k+1)}} \sqrt{w-w_0}
% \end{align}
% Which of course matches the expansion of the exact solution. 

% \section{Two-stack state}
% We can solve for the exact position of a two-stack state, \\

% \dsm{two-stack position analysis}\\

% Where LambertW is the solution 
% \dsm{definition of LambertW}\\

% Unfortunately, this isn't easy to plug in to our stability analysis. We can use an asymptotic expansion near the bifurcation point:

% \dsm{expansion}

% However this unfortunately doesn't hold for the values of $w$ where we see the next bifurcation into three stacks, since $w/w_0$ is greater than $1$.
%\dsm{show three stack unique stable solution? show four-stack branching branches? }

\subsection{Three- and four-particle cases}
\begin{figure}[ht!] 
\centering
	\includegraphics[width=.525 \columnwidth]{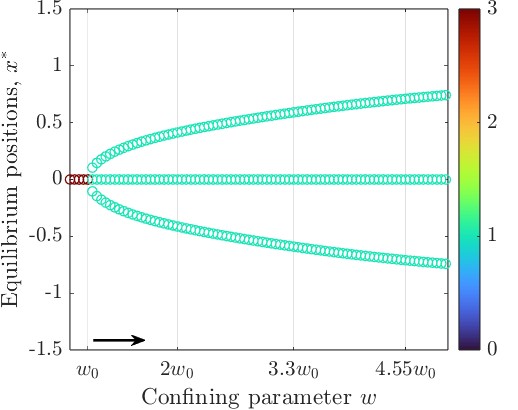}
	\includegraphics[width=.46 \columnwidth]{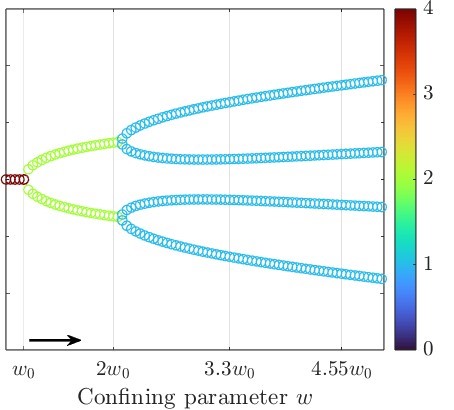}
	\caption{\textbf{Three- and four-particle bifurcation diagrams.} Equilibria for 3-particle (\textbf{left}) and 4-particle (\textbf{right}) systems. For these small $n$, these are the only stable states. Note that we have shifted to using multiples of $w_0$ on the $w$ axis, for comparison with higher $n$ cases.} 
	\label{fig:3_and_4_particles}
\end{figure}

\begin{figure}[t!] 
\centering
	\includegraphics[width=.9 \columnwidth]{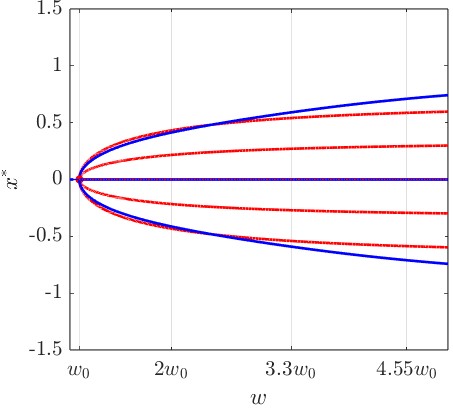}
	\caption{\textbf{$n=3$ bifurcation diagram, all equilibria.} The equilibrium diagram similar to the left panel of Fig.~\ref{fig:3_and_4_particles}, but made with analytical-continuation software (MatCont v7.3)\cite{dhooge2008new}, showing unstable equilibria (in red) as well as stable ones (in blue). We note that since this is a 2D projection of a 4D bifurcation diagram (all three state variables are superimposed on the same vertical axis), stable and unstable branches appear to ``cross'' without exchanging stability but in fact belong to entirely different branches in state space. For example, the stable curved branches correspond to the outer particles of the $1-1-1$ state, while the red branches which cross them are for the single particle in the unstable $2-1$ state (meanwhile the inner branches correspond to the location of the $2$-stack in that state). Similarly, the origin beyond $w_0$ is a stable position if the system is in the $1-1-1$ state but an unstable position for the fully-stacked state, so it is both blue and red-dotted in this figure. } 
	\label{fig:3_matcont}
\end{figure}

Three- and four-particle systems have unique stable equilibria for all $w$, which are shown in Fig.~\ref{fig:3_and_4_particles}; unfortunately these resist such easy exact-solution form as the $n=2$ case. Other equilibria exist for these systems (such as $1-2$ states\footnote{This notation indicates the partition state of the particles, i.e., the number of particles in each stack, in order. In this case, there are two visibly-distinguishable $1-2$ states since the two-particle stack can be above or below the single-particle stack, though each of those represents six equilibria in full state space due to permutations of the particles.} in the $n=3$ case, $1-2-1$ and $1-3$ states in $n=4$, and the fully-stacked origin state at $w>w_0$), but none of them are stable. In Fig.~\ref{fig:3_matcont} we show all such equilibrium positions using the MatCont analytical-continuation software package for Matlab (MatCont v7.3, see \cite{dhooge2008new}) to track those unstable artificially-partitioned states as well as the stable state we actually see in regular numerical simulations. %We can verify their uniqueness due to checking the few possible orientations in Matcont analytical-continuation software for Matlab, as pictured in Fig.~\ref{}
%\dma{Actually the 4-particle system has multiple coexistent equilibria (symmetric ones include 1-2-1 and 2-2 and all 4 stacked), but maybe not simultaneously stable.}

%\dsm{should I take $k=2$ out of the titles of all these figures for the final version?} 

The nonzero stable branches for $n=3$ (and $k=2, s=1$) are solutions to the implicit equation 
\[
 e^{8 x^{*2}} + (4 x^{*2}-4) w^2 e^{6 x^{*2}} + (32 x^{*2}-8) w^2 = 0\;,
\]
which is the result of assuming symmetry (one particle at $0$ and the other two at $x^*$) and solving for nonzero $x^*$ solutions to the equilibrium condition on the particle at $x^*$.
% \[
% 2 e^{8 x^2} + (8 x^2-8) w^2 e^{6 x^2} + (64 x^2-16) w^2 = 0\;.
% \]
For large $w \gg w_0$ this relationship converges to $w^2 = e^{2 x^2} (4-4 x^2)^{-1}$ (or explicitly in $x$: $x^2 = 1-\frac{1}{2} W(\frac{1}{2} e^2/w^2)$); for $w$ near $w_0$ ($w_0 = \sqrt{3}/6$) the relation is well approximated by $x^2 \approx 1/9 - 1/108\ w^{-2}$ instead.

%\dsm{talk more about how $2-1$ are unstable equilibriums too, and show either with implicit plot or matcont?}

%\dma{Branches in 3-particle case are solutions to this implicit equation: $2 e^{8 x^2} + (8 x^2-8) w^2 e^{6 x^2} + (64 x^2-16) w^2 = 0$, which converge to $w^2 = e^{2 x^2} (4-4 x^2)^{-1}$ (or explicitly in $x$, $x^2 = 1-\frac{1}{2} W(\frac{1}{2} e^2/w^2)$) for large $w \gg w_0$.  For $w$ near $w_0$, we have $x^2 \approx 1/9 - 1/108 w^{-2}$, with $w_0 = \sqrt{3}/6$.}
In the 4-particle case, ($n=4, k=2, s=1$) we can find the implicit equation for the $2-2$ (i.e.~two stacks of two particles each) state, namely 
\[
 w^2 = \frac{e^{8 x^{*2}}}{ 16(1 - 4 x^{*2})}\;,
\]
which is stable from $w=w_0$ ($=1/4$ in this case) to $w\approx 2.1 w_0$ as seen in Fig.~\ref{fig:3_and_4_particles}. More exactly, the $2-2$ state splits into the $1-1-1-1$ state when $w=w_c$, where
\[
w_c^2 = \frac{1 + 20 x^{*2} - 64 x^{*4}}{16(1 - 4x^{*2})}\;,
\]
and where $x^*$ satisfies the implicit equation
%or in other words at the $w$ value which yields the stack separation satisfying 
\[
1 + 20x^{*2} - 64 x^{*4} = e^{8x^{*2}}\;.
\]
%\dma{For appearance of the 1-1-1-1 state, $w_c^2 = (1 + 20 q - 64 q^2)/(16 - 64q)$, where $q \equiv x^{*2}$ satisfies $1 + 20q - 64 q^2 = e^{8q}$.}
%\dsm{how is this different from the two-stack statement above? if you simplify by substituting in for the numerator, you get the exact whole 2-2 curve}

The $1-2-1$ state, while never stable, is also tractable, with outer stack positions given by the relation
\[
8 w^2 = \frac{ e^{8 x^2}}{1 - 4 x^2 + (1-x^2)e^{6 x^2}} \;.
\]

Other particular states may also have similar implicit equilibrium expressions, but their multitude makes this endeavor an impractical strategy for understanding the system for general $n$.

\subsection{Five-particle system: birth of multistability}
%\dsm{is this worth it? it kind of jumps the gun on forward and backward passes}

\begin{figure}[t!] 
\centering
    \includegraphics[width= \columnwidth]{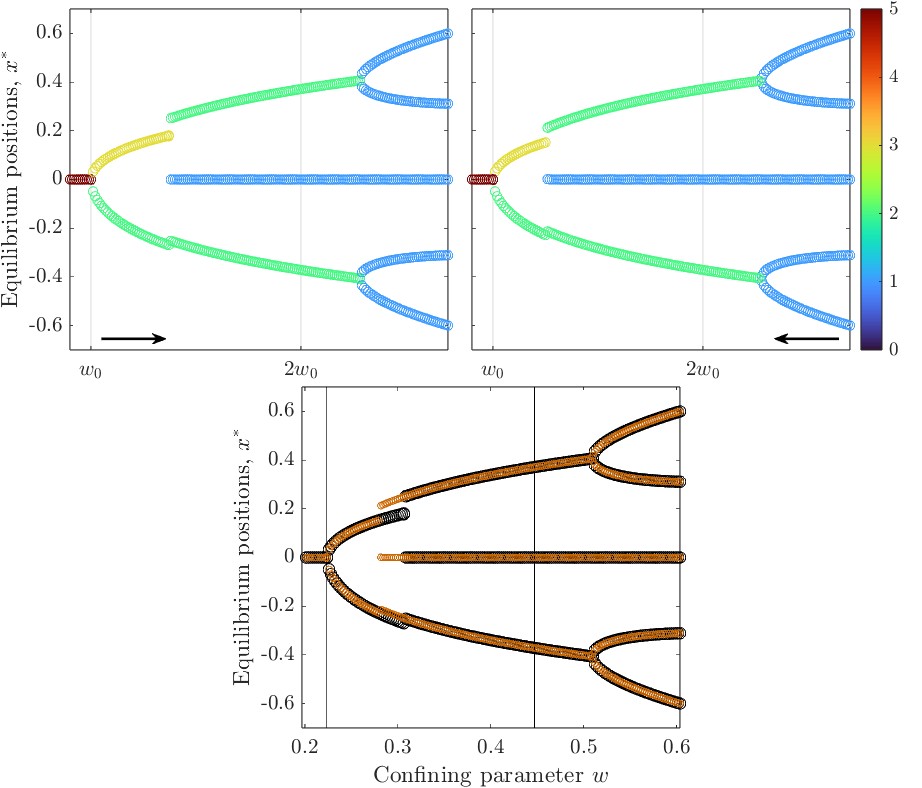}
	\caption{\textbf{Increasing- and decreasing-$w$ equilibrium diagrams, $n=5$ particles.} Zoomed equilibrium diagrams for five particles. \textbf{Top Left:} increasing-$w$ pass; note the asymmetric $3-2$ state arms are slightly longer. \textbf{Top Right:} decreasing-$w$ pass, with shorter $3-2$ arms. The overlapping region exhibits bistability of $3-2$ and $2-1-2$ states. \textbf{Bottom:} Overlay with increasing-$w$ in black and decreasing-$w$ in orange, emphasizing the area of bistability around $w=0.3$. Previous $w_0$ multiple reference points persist as vertical lines, but decimal values are provided for finer reference. 
	} 
	\label{fig:5_particles}
\end{figure}

With five particles, we see the first case where there is multistability, by two different mechanisms. First of all, unlike $n=3$ the origin cannot stably hold a particle as we cross $w_0$, and the population splits into a $3-2$ state, which is necessarily asymmetric in position. Then the indifference between which stack has $3$ particles leads to bistability between two visibly different states, though they might be considered the same state up to reflection of the domain. Second, the point that the system drops to a $2-1-2$ state (on an increasing-$w$ pass) is different from the point that it jumps back to the $3-2$ state (on a decreasing-$w$ pass). We can see this in the difference between upper-left and upper-right panels in Fig.~\ref{fig:5_particles}. This is because there is a region of bistability between $3-2$ and $2-1-2$ configurations---the loss of stability of $3-2$ happens at a higher $w$ value than the gain of stability of $2-1-2$. This hysteresis with respect to increasing and decreasing parameter $w$ is explored more generally in the ``Medium-$n$ analysis'' section below.

\subsection{Medium-$n$ analysis}
To get a sense of how the transition to the large-$n$ behavior happens, we will look at a medium-scale $n$, in particular $n=32$. 

\begin{figure}[t!] 
\centering
    \includegraphics[width= \columnwidth]{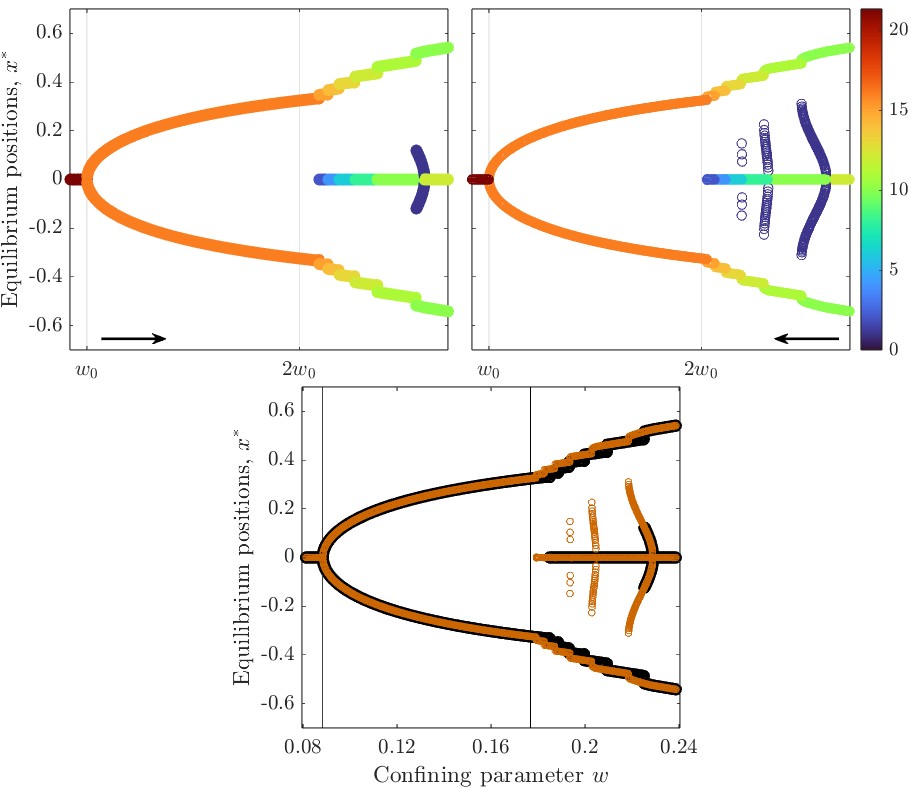}
	\caption{\textbf{Equilibrium Diagrams, $n=32$, zoomed.}  Bifurcation diagram for $32$ particles, zoomed for higher resolution, displaying particle transfer branches and the system's multistability. %, with increasing $w$ (top left), decreasing $w$ (top right), and both overlaid (bottom)
    \textbf{Top Left:} Increasing $w$ pass, where particles ``fall'' to the center only when the outer stacks lose stability at their previous capacity. \textbf{Top Right:} Decreasing $w$, with particles transferring in three visible branches when the central stack becomes ``overstuffed'' and sheds particles to the outer stacks. \textbf{Bottom:} Both passes overlaid to emphasize differences, with forward pass in black and backward pass in orange.  %At this resolution, the top right (decreasing-$w$) figure displays three branches whereby single particle pairs transfer between the central stack and the outer ones. %This simulation was done with decreasing-$w$ continuation (from right to left), so these branches show when the central stack loses stability for its over-stuffed state and sheds particles to the outer stacks as the background potential well narrows.
    } 
	\label{fig:16x2_combined}
\end{figure}

Figure \ref{fig:16x2_combined} shows part of the bifurcation diagram for $32$ particles, which shows hints of the dynamical process by which the particles transfer between stacks. At this resolution, we can see three ``connecting'' branches in the top right (decreasing-$w$) figure where single particle pairs transfer from the central stack to the outer ones.

Just as with the $n=3$ and $n=4$ diagrams, there are many more equilibria than we see in Fig.~ \ref{fig:16x2_combined}. First of all, we only see stable equilibria due to our method of forward-time numerical integration with minutely perturbed initial conditions, so we do not see the huge number of unstable equilibria%---many \dsm{can I say most? } ordered partitions of the $n$ particles, corresponding to stack sizes and their order in the potential well, have their own unstable equilibrium
. Second, we have enforced symmetry in this simulation, so we are missing the slightly asymmetric \textit{stable} states that can (and generally do) result when particles are individually free; the enforcement of symmetry is nevertheless justified as we seek a generic central pattern around which many co-stable perturbations exist. But even in the symmetric case, there is co-stability of states, which is demonstrated by the discrepancy between increasing- and decreasing-$w$ passes. Exploring these discrepancies will provide intuition about how the system behaves at higher $n$.
%we do not see in a single diagram. For example, the left panel of Fig.~\ref{fig:32_bifurc_F_and_overlay} shows the same system as Fig.~\ref{fig:32_bifurc_B} but with increasing-$w$ continuation. Exploring the discrepancies between these two will provide intuition about how the system behaves at higher $n$.

% \begin{figure}[ht!] 
% \centering
% 	\includegraphics[width=.45 \columnwidth]{32_colorscatter_F.jpg}
% 	\includegraphics[width=.35 \columnwidth]{16x2_overlay_matching.jpg}
% 	\caption{\textbf{Partial bifurcation diagram, increasing $w$ and overlay.}  \textbf{Left:} Similar to Fig.~\ref{fig:32_bifurc_B} but with increasing $w$, we see slightly different equilibrium states. \textbf{Right:} The difference between backwards and forwards passes emphasized by overlaying them, with forward pass in blue and backward pass in red. \dsm{vertical dashed lines for w0 values in overlay. put increasing pass in previous fig, and overlay on its own, or maybe combine figs entirely}} 
% 	\label{fig:32_bifurc_F_and_overlay}
% \end{figure}

 %where the rapid, particle-transfer bifurcations are too compressed to see

\subsubsection{Repeated Hysteresis}
\begin{figure}[ht!] 
\centering
	\includegraphics[width=.512\columnwidth]{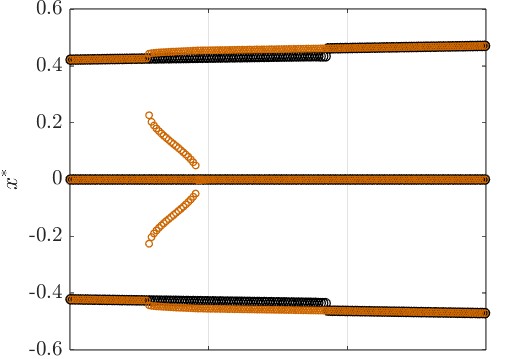}
	\includegraphics[width=.44\columnwidth]{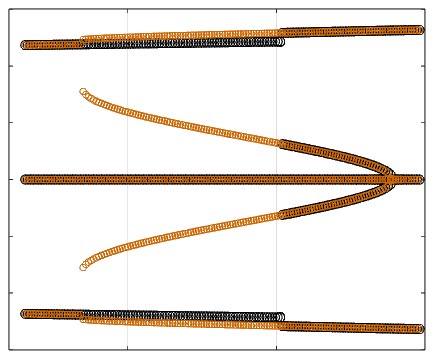}
	\\
	\begin{overpic}[width=.512\columnwidth]{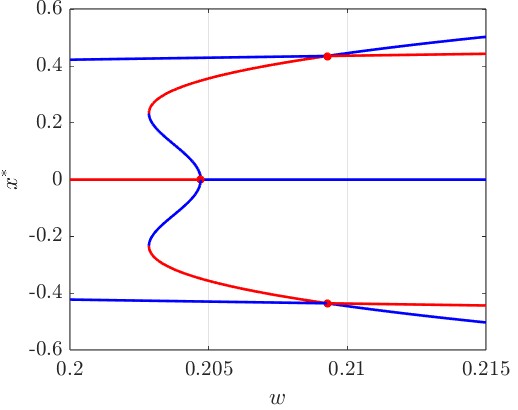}
		\put(15,70){\tiny{$12-8-12$}}
		\put(30,55){\tiny{$11-1-8-1-11$}}
		\put(15,45){\tiny{$11-10-11$}}
		\put(15,17){\tiny{$12-8-12$}}
		%\put(62,74){\tiny{$1-11-8-11-1$}} %1
		\put(50,74){\tiny{$1-11-8-11-1$}} %2
		\linethickness{1pt}
        %\put(77,73){\vector(-1.4,-1){4}} %1.1
        %\put(74,73){\vector(2.5,-1){5}} %1.2
        \put(82,74){\vector(2.5,-1){5}} %2.1
	\end{overpic}
	\begin{overpic}[width=.44\columnwidth]{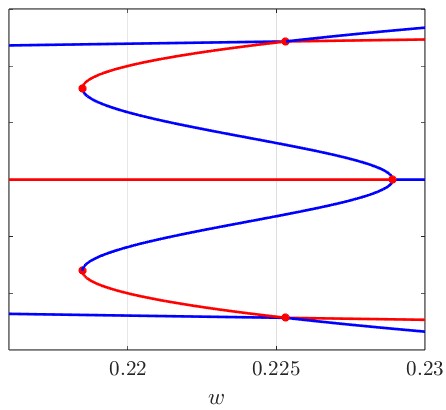}
	    \put(5,83){\tiny{$11-10-11$}}
	    \put(40,66){\tiny{$10-1-10-1-10$}}
	    \put(5,54){\tiny{$10-12-10$}}
	    \put(5,17){\tiny{$11-10-11$}}
	    %\put(54,86.5){\tiny{$1-10-10-10-1$}}
	    \put(40,86.5){\tiny{$1-10-10-10-1$}}
	    \linethickness{1pt}
	    \put(79,87.5){\vector(2.5,-1){5.5}}
	\end{overpic}

	\caption{\textbf{Particle-transfer branches.}  Zoom views of central and right ``transfer arms'' from Fig.~\ref{fig:16x2_combined}.  \textbf{Top:} Overlaid scatter plots of the full-population equilibria, for forward (black) and backward (orange) passes. \textbf{Bottom:} Stability of states evaluated using MatCont analytical-continuation software, tracking locations where the transferring ``free'' particle pair can be---with stable positions are shown in blue, and unstable in red. The free pair may align with the other stacks (seen as effectively-horizontal lines, like the state labeled $12-8-12$; these persist through bifurcations but switch stability), or may reside in-between or outside them (e.g.,~the $11-1-8-1-11$ state which becomes $1-11-8-11-1$ in the bottom-left figure).   The left figures explore %the hysteresis involved in 
	the empirically observed transition from a $12-8-12$ state on the low $w$ side to $11-10-11$ for higher $w$, and the right figures explore the transition from $11-10-11$ (lower $w$) to $10-12-10$ (higher $w$). %Population states caused by that pair's position are labeled in our ``partition'' notation; the horizontal states like $12-8-12$ and $11-10-11$ persist through bifurcation points but switch stability, while the curved states representing unaligned transfer-pair positions cross the outer states and are renamed due to the notation (like $11-1-8-1-11$ becoming $1-11-8-11-1$ in the bottom-left figure).
	For these parameters, $w_0 = 0.0884$. } 
	%old caption (fits)
% 	\caption{\textbf{Particle-transfer branches.}  Zoom of hysteresis for particle transferring between middle and outer stacks. \textbf{Top:} Overlaid scatter plots of the full-population equilibria. Stack-size information is omitted in favor of clarifying forward (black) and backward (orange) passes. \textbf{Bottom:} Stability of states from MatCont analytical-continuation software, tracking places where the ``transferring'' particle pair can be---aligning with the outer stacks the central stack, in between, or even outside the outer stack.  Stable positions are shown in blue, and unstable in red. The left figures explore the transition from a $12-8-12$ state on the low $w$ side to $11-10-11$ for higher $w$, and the right figures explore the transition from $11-10-11$ (lower $w$) to $10-12-10$ (higher $w$). For these parameters, $w_0 = 0.0884$. } 
	\label{fig:32_branch_zooms}
\end{figure}
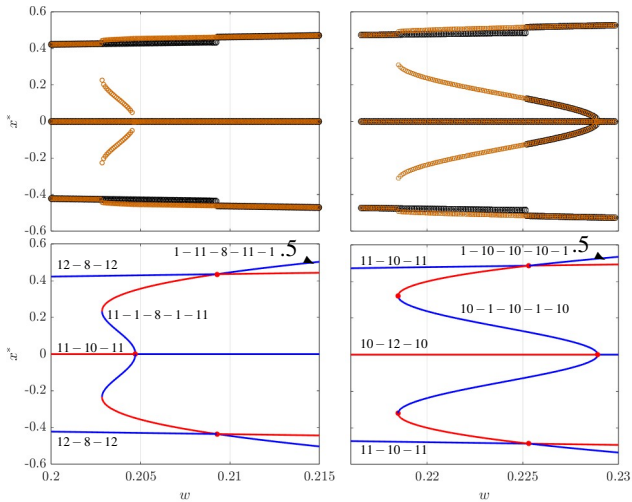

The bottom panel of \ref{fig:16x2_combined} emphasizes the differences between the increasing- and decreasing-$w$ passes, with increasing-$w$ pass in black and decreasing-$w$ in orange. Two of those ``branches'' are isolated in Fig.~\ref{fig:32_branch_zooms}.

To understand the hysteresis in Fig.~\ref{fig:32_branch_zooms}, we start at the left side of the left-two figures, at $w=0.2$. The top-left figure, a scatter plot of the full population, hides the stack-size information, but this is a $12-8-12$ arrangement in both black (increasing-$w$) and orange (decreasing-$w$) passes. The bottom-left figure shows us why; the ``free'' pair of particles may only stably exist at the outer stack positions for this parameter value (the central position is red, indicating instability). 

As the parameter value increases, this arrangement stays stable until slightly under $w=0.21$, when we see this branch undergoes a transcritical bifurcation. Theoretically, the pair could drift outward beyond the outer stack at this point, as the lower-left figure indicates, but that diagram assumes the other stacks stay perfectly stacked, while in reality we perturb all particles with noise, and that state isn't robust to that broken symmetry. So what we actually see is that one pair falls to the origin and the rest remain together, corresponding to the $x^*=0$ stable state in the bottom-left figure, and the overall population state $11-10-11$. 

However, as we decrease $w$ again, the system stays at this $11-10-11$ state until the branch point near $w=.205$, where the free pair's position follows stable branches away from the origin (in a $11-1-8-1-11$ state) until those branches go vertical in an apparent saddle-node bifurcation, at which point the pair jump suddenly to join the outer stacks again. During this bistable region, the position of the outer stacks differs slightly, which is reflected in the disalignment of outer stacks in the top-left figure.

A similar process occurs at slightly higher parameter value, reflected by the right two figures. In this case, the population is transitioning between $11-10-11$ and $10-12-10$ states. The only qualitative difference this time is the transcritical loss of stability at the outer position occurs \textit{before} the central pitchfork bifurcation, so the outer pair is dropped to that branch of the pitchfork in a $10-1-10-1-10$ state on the forward pass rather than all the way to the center. On the backwards pass, of course, that $10-1-10-1-10$ state persists longer before losing stability and the transferring particles rejoin the outer stacks.

In this way, we see how bistability occurs between distinct population fractionations. As we can see in Fig.~\ref{fig:16x2_combined}, the fractionation changes more rapidly when the central stack is small, which we may now understand causes these branches to overlap, yielding multi-stability between more than two fractionation states. Furthermore, as Fig.~\ref{fig:colorscatter_convergence} displays, these transitionary states become narrower (in $w$) as $n$ increases, such that we no longer easily see them at finite resolution. In the infinite-$n$ limit, there is a continuous family of these bifurcations (and corresponding family of transition curves) as the central fractionation changes smoothly rather than in these discrete jumps, and smooth bands of stable fractionation (and corresponding stack positions) accordingly.

% \subsection{Large-$n$ analysis}
% The really interesting results occur with large numbers of particles. 

% We can check the stability of particular configurations using continuation software, for example testing the stable and unstable equilibrium positions of a $129^\textrm{th}$ ``test" particle given $128$ particles in two stacks of $64$:

% \dsm{figure: matcont boat?}

% We can see the test particle can align with either of the large stacks (desymmetrizing their locations imperceptibly). But we also see that the birth of stability at the origin is in fact due to a second pitchfork bifurcation with very short-lived asymmetric unstable branches which cross the outer stack positions and become stable, roughly corresponding to sitting in the well outside the two large stacks (smoothly approaching that well location, $3s/2$, as $w to \infty$).

\subsection{Large-$n$ analysis}
\begin{figure}[t!] 
\centering
    \includegraphics[width=\columnwidth]{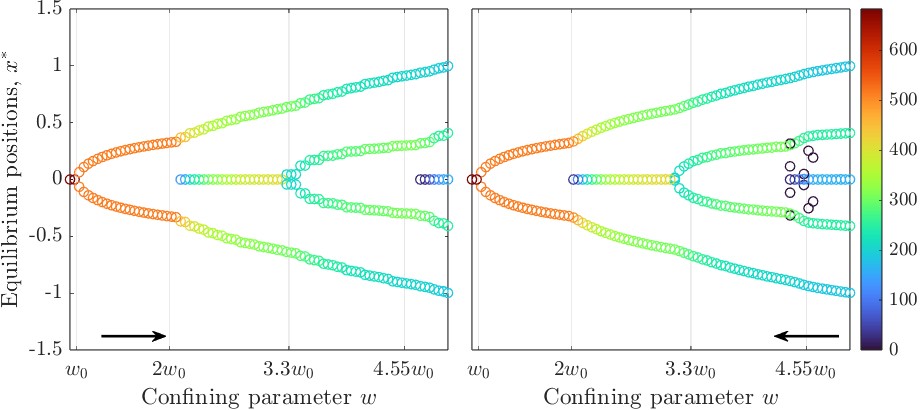}
	\caption{\textbf{Equilibrium diagrams, large-$n$ forward and backward passes.} Equilibrium diagrams for $1024$ particles, with symmetry enforced. Some differences between forward and backward passes indicate the persistence of stability bands: we see different stable fractions of the population at the origin, and correspondingly different bifurcation points of the origin stack.} 
	\label{fig:512x2_FB_comparison}
\end{figure}

As Fig.~\ref{fig:colorscatter_convergence} suggests, the overall system behavior appears to converge for large numbers of particles, under the appropriate scaling of the $w$ axis. This makes discussion of the large-$n$ limit meaningful---indeed it appears that the rapid ``transitionary'' bifurcations from the previous section become effectively invisible, while the ``major'' central stack-birth/stack-splitting bifurcations remain. There is still fractionation indifference (i.e.~bands of possible stable population percentages in each stack), and the location of these major bifurcations can still vary meaningfully between forward and backward parameter continuation, as Fig.~\ref{fig:512x2_FB_comparison} shows.

\begin{figure}[t!] 
\centering
	\includegraphics[width=.9\columnwidth]{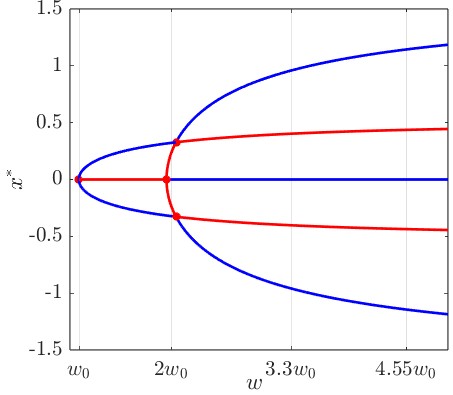}
	\caption{\textbf{Stable and unstable equilibrium positions for a $129^{\textrm{th}}$ particle.} Analytical-continuation software Matcont yields the stable (blue) and unstable (red) equilibrium positions for a $129^{\textrm{th}}$ particle in a system with two perfectly-aligned stacks of $64$ particles (which reside at the narrower U-shape, technically but imperceptibly influenced by the position of this $129^{\textrm{th}}$ particle). We see that the ``test particle'' can stably align with those two stacks from $w = w_0$ up to slightly above $w = 2 w_0$, the latter point occurring right after the birth of stability at the origin. At that point, it must either fall to the center or flee to the outside trough position. In this context, we see stability at the center is born as a pitchfork bifurcation, and we may observe %clearly see that there is % As $n$ increases, that pitchfork gets more abrupt/``square'', but there is still 
	the small region of bistability shared between $64-65$ and $64-1-64$ states (which includes $2 w_0$). For $n=129$, $w_0=0.0442$.} 
	\label{fig:129_matcont}
\end{figure}

We can check the stability of particular configurations like we did in Fig.~\ref{fig:3_matcont} for $3$ particles, again using MatCont---for example, testing the stable and unstable equilibrium positions of a $129^\textrm{th}$ ``test'' particle given $128$ particles in two stacks of $64$---this is shown in Fig.~\ref{fig:129_matcont}.

In Fig.~\ref{fig:129_matcont} we see that the test particle can align with either of the large stacks (desymmetrizing their locations imperceptibly). But we also see that the birth of stability at the origin is in fact due to a second pitchfork bifurcation with very short-lived asymmetric unstable branches which cross the outer stack positions and become stable, roughly corresponding to sitting in the trough outside the two large stacks (and in fact approaching that well location, $\pm 3s/2$, as $w \to \infty$).

\section{Analysis}
\subsection{General Equilibrium Statement}
For particle $j$ to be at equilibrium:
\begin{align}
    &\deriv{x_j}{t} = -\deriv{U}{x}\biggr\rvert_{x_j} = 0, \nonumber\\
    &0= -\frac{2x_j}{w^2} + \sum_{i=1}^n \frac{2k^2(x_j-x_i)}{(k-1)s^2}\left( 1 - \frac{(x_j-x_i)^2}{s^2} \right) e^{-k\frac{(x_j-x_i)^2}{s^2}}\;. \label{eq:gen_equil}
\end{align}
If the right hand side (RHS) is positive, particle $j$ will move right, and if negative, left. If it is zero for all particles, the system is at equilibrium. There are many particular states, such as all the particles at the origin $(0,0,\ldots,0)$, two symmetric stacks $(x^*,x^*,\ldots,x^*,-x^*,-x^*,\ldots,-x^*)$, etc., which may satisfy this equilibrium condition for various parameter values.\\

\subsection{Stability} \label{sec:general-n_stability}
We can analytically examine the stability of the fully-stacked state at the origin, recovering the critical bifurcation value $w_0$ below which that state is stable---and in fact appears to be the \textit{only} equilibrium.\\

The elements of the $j^{\textrm{th}}$ row of the Jacobian matrix $J$ for the system \eqref{eq:gen_equil}  are 
\begin{equation}
  \left\{\begin{aligned}
 J_{jj} &= -\frac{2}{w^2} + \frac{2k^2}{(k-1)s^2}\cdot\\
  & \sum_{i=1}^n\left[ 1 - \frac{3+2k}{s^2}(x_j-x_i)^2  + \frac{2k}{s^4}(x_j-x_i)^4  \right] e^{-\frac{k(x_j-x_i)^2}{s^2}} \\
    J_{ji} &= -\frac{2k^2}{(k-1)s^2} \cdot \\
    &\left[ 1 - \frac{3+2k}{s^2}(x_j-x_i)^2  + \frac{2k}{s^4}(x_j-x_i)^4  \right] e^{-\frac{k(x_j-x_i)^2}{s^2}}
\end{aligned}\right.  \;,
\label{eq:Jacobian_general}
\end{equation}
% \begin{align}
%     J_{jj} = -\frac{2}{w^2} + \frac{2k^2}{(k-1)s^2}\sum_{i=1}^n\left[ 1 - \frac{3+2k}{s^2}(x_j-x_i)^2  + \frac{2k}{s^4}(x_j-x_i)^4  \right] e^{-\frac{k(x_j-x_i)^2}{s^2}} \\
%     J_{ji} = -\frac{2k^2}{(k-1)s^2}\left[ 1 - \frac{3+2k}{s^2}(x_j-x_i)^2  + \frac{2k}{s^4}(x_j-x_i)^4  \right] e^{-\frac{k(x_j-x_i)^2}{s^2}}
% \end{align}
which, at the origin $x_i = x_j = 0$ (corresponding to the fully-stacked state), become
\begin{align*}
    J_{jj}|_O &= -\frac{2}{w^2} + (n-1)\frac{2k^2}{(k-1)s^2}\;, \\
    J_{ji}|_O &= -\frac{2k^2}{(k-1)s^2}\;. 
\end{align*}
Due to the symmetry, we can identify all the eigenvectors: 
\begin{enumerate}
    \item The eigenvector $v_1 = (1,1,\ldots, 1)$ corresponding to the full stack drifting left or right from the origin has eigenvalue $\lambda_1 = -2/w^2$, which is always negative, indicating that the 1-stack system is stable to these types of perturbations (unsurprising based on intuition for a single particle). 
    \item The other $n-1$ eigenvectors consist solely of symmetric two-particle divergence; i.e., vectors of the form $(-1,1,0,0,\ldots,0)$ with the positive $1$ in each of the other $n-1$ positions. These vectors all have eigenvalue $\lambda = -2/w^2 +2nk^2/(k-1)s^2$. The 1-stack state is thus stable to these types of perturbations for
    \begin{align}
        \frac{w}{s}&< \frac{1}{k}\sqrt{\frac{k-1}{n}} \;. \label{eq:global_collapse}
    \end{align}
    %which indicates the relationship for which the fully-stacked origin state is stable. 
    Solving for $w$, this gives us the critical parameter value $w_0$ for general $n$:
    \begin{equation}
        w_0= \frac{s}{k}\sqrt{\frac{k-1}{n}}\;. \label{eq:w0_general}
    \end{equation}
    We note that this agrees with the $n=2$ particular case, given above after Eq.~\eqref{eq:xapprox}. This also suggests the appropriate way to rescale the $w$ axis as $n$ varies (as we have for all figures), since the structure appears to depend only on the ratio $w/w_0$ (at least asymptotically for large $n$). %This also gives our $w$-axis scaling, $w \sim n^{-1/2}$, since the structure is dependent on multiples of $w_0$. \dma{I'd say ``This also suggests a way to rescale the $w$ axis as $n$ varies, since the structure appears to be (at least asymptotically for large $n$) dependent only on the ratio $w/w_0$.''}
\end{enumerate}

\subsection{Two-stack state}
The simplest nontrivial arbitrary-$n$ case, the two-\textit{stack} state---where the population is split into two symmetric stacks of $n/2$ particles each---is quite relevant to examine since it is the dominant behavior for approximately $w_0<w<2w_0$. It is exactly solvable via a simple tweak of the logic which led us to Eq.~\eqref{eq:2_particle_equil}, with the influence from the other ``particle'' being multiplied by $n/2$ while the background contribution is unchanged: 
\begin{align*}
    \left.\deriv{U_{tot}}{x}\right|_{x^*} = 0 &= \frac{2x^*}{w^2} - \left(\frac{n}{2}\right)\frac{4k^2x^*}{(k-1)s^2}\left( 1 - 4\frac{x^{*2}}{s^2} \right) e^{-\frac{4kx^{*2}}{s^2}} \\
     %0 &= 2x^* \left[ \frac{1}{w^2} - \frac{nk^2}{(k-1)s^2}\left( 1 - 4\frac{x^{*2}}{s^2} \right) e^{-\frac{4kx^{*2}}{s^2}}\right]\\
     \implies x^* &=0 \qquad \textrm{or} \\
     w^2 &= \frac{s^2 (k-1)}{n k^2} \frac{1} {1-4\frac{x^{*2}}{s^2}}e^{\frac{4kx^{*2}}{s^2}} \\
     w^2 &=  w_0^2 \frac{e^{\frac{4kx^{*2}}{s^2}}} {1-4\frac{x^{*2}}{s^2}}\numberthis \label{eq:2_sym_stacks_equil}
\end{align*}

%\dsm{general sol:}
The solution can be written explicitly in terms of the Lambert W function (as done above for the special case $n=2$), now using $w_0$ to further simplify:
% \begin{equation}
%     x^*(w) = \frac{s}{2} \sqrt{1 - \frac{1}{k}W\left(\frac{(k-1)}{k}\frac{s^2 e^k}{n w^2}\right)}.
% \end{equation}
%
\begin{equation}
  x^*(w) = \frac{s}{2} \sqrt{1 - \frac{1}{k}W\left(k e^k\frac{w_0^2}{w^2}\right)} 
 \end{equation}

For $|x| \ll 1$, this is approximately
 \[
   x^* \approx \frac{s}{\sqrt{2w_0(k+1)}} \sqrt{w-w_0}\;.
 \]
exactly as we saw before (all scaling is accounted for by $w_0$).

\subsection{Return of stability at the origin}
We seek an understanding of the second ``major'' bifurcation: the birth of the three-stack state near $w=2w_0$, with a small central stack between the two large symmetric stacks. We can easily check the curvature of the potential landscape at the origin between two equal stacks, using this as a test to identify when a particle would stably rest there. We note that this spot sees an identical contribution from all particles at $\pm x^*$:
\begin{align*}
    \left.\frac{\textrm{d}^2U_{tot}}{\dd{x}^2}\right|_{x=0} &= \frac{2}{w^2} + n\left[-\frac{2k^2}{(k-1)s^2}\cdot \right.\\
    &\qquad \left. \left( 1-\frac{2k+3}{s^2}x^{*2}+\frac{2k}{s^4}x^{*4}\right)e^{-\frac{kx^{*2}}{s^2}} \right]\\
    &= \frac{2}{w^2} - \frac{2}{w_0^2}\left( 1-\frac{2k+3}{s^2}x^{*2}+\frac{2k}{s^4}x^{*4}\right)e^{-\frac{kx^{*2}}{s^2}} 
    ;,
\end{align*}
so the birth of stability happens when this curvature crosses $0$, at
\begin{equation}
    w_c^2 = w_0^2\frac{e^{\frac{kx^{*2}}{s^2}}}{1-\frac{2k+3}{s^2}x^{*2}+\frac{2k}{s^4}x^{*4}} \label{eq:w_c_implicit}\;.
\end{equation}
When we combine this condition with the two-stack equilibrium relation for $w^2$, Eq.~\eqref{eq:2_sym_stacks_equil}, we get 
\begin{align*}
    w_0^2\frac{e^{\frac{kx_c^{*2}}{s^2}}}{1-\frac{2k+3}{s^2}x_c^{*2}+\frac{2k}{s^4}x_c^{*4}} 
    &= w_0^2 \frac{e^{\frac{4kx_c^{*2}}{s^2}}} {1-4\frac{x_c^{*2}}{s^2}}\;, \\
    1-4\frac{x_c^{*2}}{s^2} &= \left(1-\frac{2k+3}{s^2}x_c^{*2}+\frac{2k}{s^4}x_c^{*4}\right) e^{\frac{3kx_c^{*2}}{s^2}} \numberthis \label{eq:x*_for_origin_stability}\;,
\end{align*}
which defies closed-form solution but which we may numerically approximate for our default parameters $k=2$ and $s=1$, yielding $ x_c^{*} \approx 0.324$. We can see that this agrees empirically with the stack width coincident with the birth of stability in our large-$n$ figures like Figs.~\ref{fig:256x2_F_w_3D} and \ref{fig:512x2_FB_comparison}. 

This approximation for $x^*$ in turn allows us to approximate $w_c$, by using either relation again. Using Eq.~\eqref{eq:2_sym_stacks_equil} with $k=2, s=1$, we have
\begin{align*}
     w_c^2 &= w_0^2 \frac{e^{8x_c^{*2}}} {1-4x_c^{*2}} 
     \;,\\
    x_c^* &\approx 0.324176\ldots \\
   \implies w_c & \approx 1.99978 w_0 \;.
\end{align*}
This is suspiciously close to $2 w_0$, but these approximations were done using $16$-digit precision, so it appears to indeed be distinct. We note that this value depends on $k$ (though $s$ may be scaled out as always); for example, for $k=100$ we have $w_{c,100} \approx 1.791 w_0$.

After the central stack's creation, the exchange of particles between central and outer stacks is complicated, since the fraction of the population at the origin influences the position of the outer stacks, which in turn influences the stable fraction at the origin. We observe empirically that the outer stacks drift apart in a roughly linear manner, which may be helpful, though we leave this exploration for future work. 

We note that the increasing-$w$ sweep only sees % the \textit{minimum} stable amount of 
particles at the center %, since particles must be
when they are kicked out of the outer stacks to populate it. On the decreasing-$w$ pass, however, the center hosts a larger stable population at each $w$ value reached, since continuation from the right causes the accumulation of all nearby particles at the center, where they stay until they are ejected to the outer stacks. It is perhaps counterintuitive that these particles ``climb'' the global potential as it narrows, but it is nevertheless true; the narrowing background potential pushes the outer stacks inward enough that the center becomes less stable, at which point it is ``overstuffed'' and repels some of its former constituents to join the other stacks.  

\begin{figure}[ht!] 
\centering
	\includegraphics[width=.9\columnwidth]{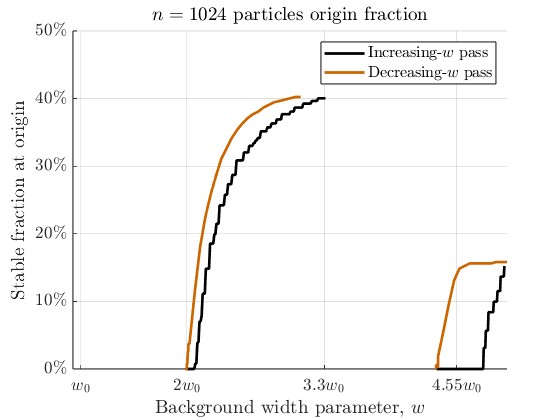}
	\caption{\textbf{Origin fractionation bands, $n=1024$.} Fraction of population at origin for forward and backward passes with small noise. These curves demonstrate the existence of bands of stability, and are an approximation of those bands. The true bands may be slightly wider, but the fractionations between these curves are stable for corresponding $w$ values. 
	 %Any fractions between these curves correspond to a stable state as well, but this family of states is difficult to fully characterize since near-origin behavior is interdependent with outer stack populations and positions.%These curves may act as upper and lower bounds for stable fractionation at the origin. 
	} 
	\label{fig:512x2_FB_origin_fraction_comparison}
\end{figure}

The decreasing pass thus acts as a lower bound for the \textit{maximal} stable fraction at the center, and the increasing pass acts as an upper bound on the \textit{minimal} stable fractionation. We expect a continuous band of stable fractionations between those values, as indicated by Fig.~\ref{fig:512x2_FB_origin_fraction_comparison}. We leave further exploration of the bands of stability in the $n \to \infty$ limit for future work. %\dsm{tweak this if actually do the full hysteresis cycle}

%and for large $n$ the fraction of the population in the middle can be approximately \dma{Finish?  Maybe mention expectation of bands of stability here.}

 \subsection{$k$ dependence} \label{sec:Ricker_k_dependence}
\begin{figure*}[t!] 
\centering
    \includegraphics[width=.48\textwidth]{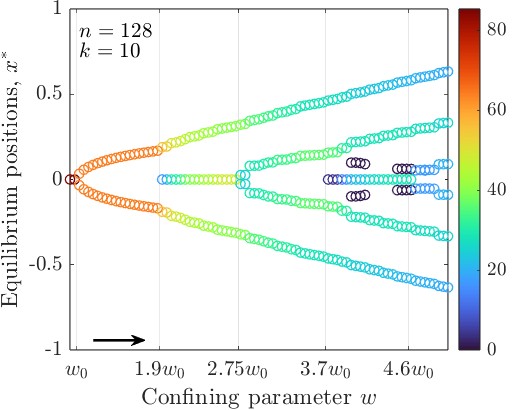}
    \includegraphics[width=.48\textwidth]{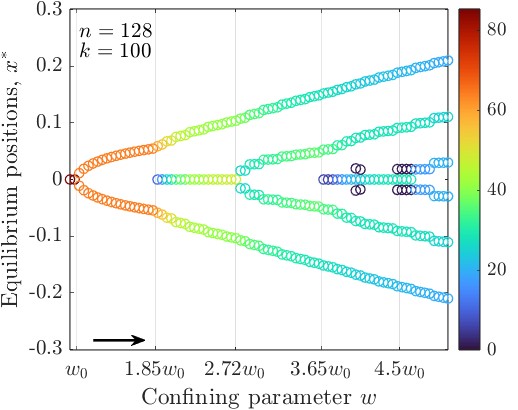}
	\caption{\textbf{$k$ dependence.} Equilibrium diagrams for higher values of $k$. The estimated bifurcation values are marked on the $w$ axis. As well as narrowing the system in $x$, larger $k$ appears to accelerate the bifurcation rate in $w$, but with diminishing returns, hinting at the invariant limit we find in Section \ref{sec:Ricker_k_dependence}: $x_c^* \to \approx 0.2914$ and $w_c \to 1.791w_0$ as $k \to \infty$. %\dma{possibly to be corrected}
    } 
	\label{fig:k10_k100}
\end{figure*}

The parameter $k$ distorts the equilibrium diagram; in particular, Fig.~\ref{fig:k10_k100} shows what happens when $k$ grows. We see that large $k$ coincides with a narrower population (smaller range on vertical axis); this is perhaps unintuitive since large $k$ makes the Ricker wavelet's trough shallower, which intuitively means ``less stability'' overall, and perhaps the dominance of repulsion. However, that would imply a wider population, while in reality we see a narrower one. We might instead take away the competing explanation that the Ricker wavelets' weaker attraction to their preferred distance means the particles don't ``buoy'' their partners away from the origin as strongly, which leads to more dominance of the background potential-well containment overall.

Asymptotic analysis of Eqs.\ref{eq:w_c_implicit} and \ref{eq:x*_for_origin_stability} with large $k$ yields 
\begin{align*}
    x_c &\sim \frac{\alpha}{\sqrt{k}}, \textrm{ where} \\
    &(1-2\alpha)e^{3\alpha}=1 \\
    \implies\alpha&=\frac{1}{3} W\left(-\frac{3}{2} e^{-3/2}\right) + \frac{1}{2} \\&\approx 0.2914 , \textrm{ and}\\
    w_c  &\to w_0 e^{2 \alpha} \approx 1.791 w_0.
\end{align*}
which seems compatible with the large-k diagrams shown in Fig.~\ref{fig:k10_k100}.
%$x_c \sim \frac{\alpha}{\sqrt{k}}$ where $\alpha=\frac{1}{3} W(-\frac{3}{2} e^{-3/2}) + \frac{1}{2} \approx 0.2914$ (that is, $\alpha$ solves $(1-2\alpha)e^{3\alpha}=1$)---narrower with larger $k$---and $w_c  \to w_0 e^{2 \alpha} \approx 1.791 w_0$. %\dsm{should I include any of this work? It's $k = 1/\epsilon^2, x \sim c \epsilon$ to balance, Eq. \ref{eq:x*_for_origin_stability} gives $c= s/\sqrt{6}$, then Eq.\ref{eq:w_c_implicit} for $w_c$}
%\dma{I agree with the method but I get different results.  I find $x_c \sim \sqrt{\alpha/k}$, where $\alpha=\frac{1}{3} W(-\frac{3}{2} e^{-3/2}) + \frac{1}{2} \approx 0.2914$ (that is, $\alpha$ solves $(1-2\alpha)e^{3\alpha}=1$). I then get that $w_c  \to w_0 e^{2 \alpha}$.  Numerically this is close but different: mine is $w_c \to 1.791 w_0$, yours is $w_c \to 1.772 w_0$.  If I leave $s$ in the problem, I get  $x_c \sim s \sqrt{\alpha/k}$ and still $w_c \to w_0 e^{2 \alpha}$. So for large $k$ there is no $s$-dependence in the problem, even if $s$ is not removed via rescaling parameters.}

%\dma{What about $k \to 1^+$?  From what I see in Eq.~16 and Eq.~13 it seems like nothing special happens there, strangely.  Just $x_c \approx 0.43$ and $w_c \approx 2.74 w_0$.  But on physical grounds it seems like something special really should happen there.  Maybe the calc for the appearance of the 3-stack state is irrelevant for other reasons?  Perhaps the 2 stack state never exists as $k \to 1^+$? But from Eq.~12 the single stack should still lose stability for small nonzero $w$.  So I'm puzzled.}

\section{Methods/Simulation details} \label{sec:methods}
%\dsm{should this go before analysis?} 
Particles were started with random Gaussian positions with standard deviation $0.1$ on the small-$w$ end. The positions were updated using ODE45 numerical integration for an initial duration of $T=20 n^{-1/2}$ time units (chosen empirically to approximate equilibration-time scaling with $n$). Runs at each $w$ value exponentially scaled integration time until all particles had moved less than $0.001$ units, or a run ended with $T>5000n^{-1/2}$ (which occured after $8$ doublings, for a maximum total run-time of $10,220 n^{-1/2}$ time units for any single $w$ value). This was necessary since equilibration time grows dramatically near bifurcation points. 

After each integration converged or hit the time limit at one $w$ value, final positions were recorded and a small random perturbation of each particle's position (Gaussian with standard deviation $0.001$) was applied to that ending state before the parameter value $w$ was updated and the next simulation commenced---this ensured we only recorded stable equilibria. This process proceeded from $w = 0.925 w_0$ to $5 w_0$ before descending along those same values; the black arrows in each diagram indicates the direction of this continuation. 

\section{Discussion/Conclusion}
We have examined a system of particles with first-order coupling through Ricker wavelet potential functions, and found remarkably rich self-organizing behavior. Intuitive small-$n$ cases transition to archetypal large-$n$ limiting behavior, with non-origin bifurcations becoming compacted into invisibility while other, ``major'' bifurcations (those regarding stability of the origin) persist and stabilize for large $n$. Multi-stability abounds, and persists for large $n$; overlapping hysteresis in the position of individual particles becomes stability bands for fractions of the population.

There is plenty more to be explored with these particles. We have not systematically examined dependence on the parameter $k$, which controls trough depth, though we showcase some promising initial findings in section \ref{sec:Ricker_k_dependence} in the appendix below. Also of interest is the oscillator interpretation, with these particles living on a finite periodic domain, for which we likewise present some intriguing initial findings in the appendix. Behavior with other types of confinement, such as a finite ``hard-walled'' box, is also an open question. Some of these model variants may lend themselves to real-world applications such as those mentioned in the introduction. 

We hope this work provides a solid foundation for the exploration of this system and its variants, which offer tantalizing glimpses of order governing a wild dynamical zoo of possible behavior.

\appendix
\section{Extra Findings}

\subsection{Fresh random starts}
\begin{figure}[t!] 
\centering
	\includegraphics[width=.9\columnwidth]{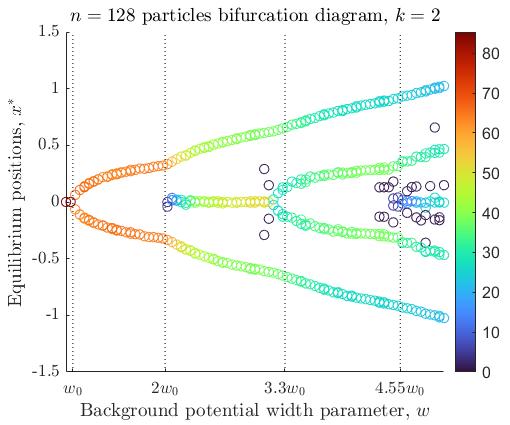}
	\caption{\textbf{Equilibrium diagram without continuation.} Equilibrium states with a fresh Gaussian random start every time (and symmetry not enforced). Stack size here gives a lower bound for tolerance of asymmetric fractionation; the greatest asymmetry in the two-stack state observed here is $68-60$.} 
	\label{fig:128_random_start_every_time}
\end{figure}

Without continuation---i.e.,~when the simulation at each $w$ value proceeds from an entirely random starting state rather than a slightly-perturbed version of the previous equilibrium---we see a ``fuzzier'' but ultimately similar picture; see Fig.~\ref{fig:128_random_start_every_time}. This does yield some information about the system's tolerance for asymmetry; the two-stack state for $n=128$ can be as lopsided as $68-60$ in this run and still appear stable. More lopsided stable two-stack states may be possible with biased initial conditions, however. %\dsm{though this might just be due to the starting states being mostly balanced... worth directly investigating or not?}

% \subsection{$n=5$ birth of multistability}

% \dma{Also figure / comments about bands of stability for large $n$ (or $n \to \infty$).  And maybe comment on odd vs even $n$?}

\subsection{Lennard-Jones and Morse potentials} \label{sec:LJ_and_Morse}
As mentioned in Section \ref{sec:intro}, the Lennard-Jones and Morse potentials from physics are models of intermolecular potential energy with short-range repulsion and long-range attraction. These potentials may be described in the following forms:
\begin{align*}
    U_{\textrm{LJ}}(x) &= 4 \epsilon \left[\left(\frac{x}{\sigma}\right)^{-12} -\left(\frac{x}{\sigma}\right)^{-6} \right] \;,\\
    U_{\textrm{M}}(x) &= D_e \left[e^{-2 a (|x|-r_e)} - 2 e^{-a (|x|-r_e)} \right] \;.
\end{align*}

\begin{figure}[ht!] 
\centering

    \includegraphics[width=.48\columnwidth]{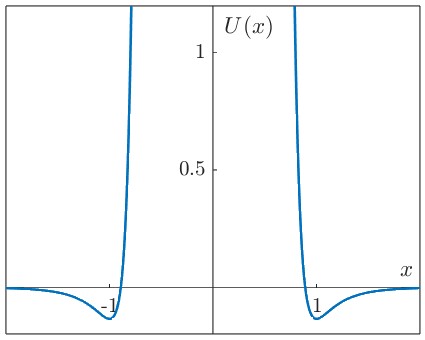}
    \includegraphics[width=.48\columnwidth]{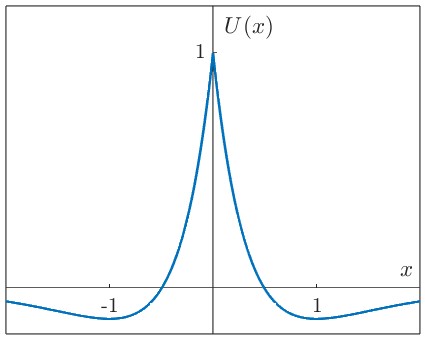}
	\caption{\textbf{Other Classic Potentials.} Examples of the Lennard-Jones (\textbf{left}) and Morse (\textbf{right}) potentials from physics. Parameters have been chosen to match the trough coordinates of our default-parameter Ricker potential---$(\pm 1, e^{-2})$---and the peak of $(0,1)$ for the Morse potential. Still, we note significant qualitative discrepancies, namely the infinite central spike for Lennard-Jones and the ``sharp'' origin peak for Morse. These qualities preclude the stability (or even well-defined behavior) of stacking behavior, and thus the particular richness of behavior we find in our Ricker system, but alterations to smoothen behavior at the origin may lead to reconciliation.  } 
	\label{fig:LJ_and_Morse_potentials}
\end{figure}

Fig.~\ref{fig:LJ_and_Morse_potentials} shows examples of these potential functions, with as much matching to our default Ricker potential as possible. In particular, this should highlight the limits of their qualitative comparability, and why alteration to a ``soft-core'' potential amenable to ``stacking''/coexistence would be necessary to expect results similar to those presented in this work.

\subsection{Ricker Oscillators} \label{sec:Oscillators}
A model variant of considerable interest for these Ricker-potential-coupled particles is their implementation as coupled oscillators. In this case, their position would represent phase on a periodic domain, like $(-\pi,\pi]$. A slight tweak to the Ricker potential would need to be defined to make it periodic; distance between particles in this space might be taken to be the shortest distance around the circle, or the infinite sum of possible distance interpretations at all $\pm 2\pi m$ multiples, or the potential itself might be made periodic in some other way. In any case, the parameter $s$ (controlling the location of the troughs, which acts as a ``preferred distance'') is no longer removable by scaling in this paradigm; its ratio with the domain is a qualitatively important value.

\begin{figure*}[ht!] 
\centering
	\includegraphics[width=.32\textwidth]{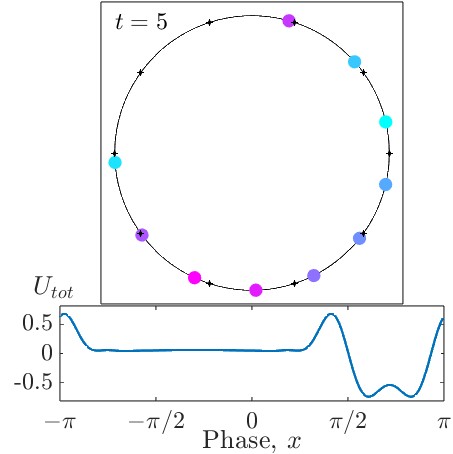}
	\includegraphics[width=.32\textwidth]{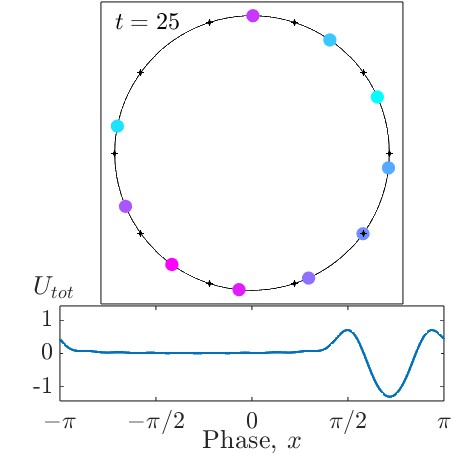}
	\includegraphics[width=.32\textwidth]{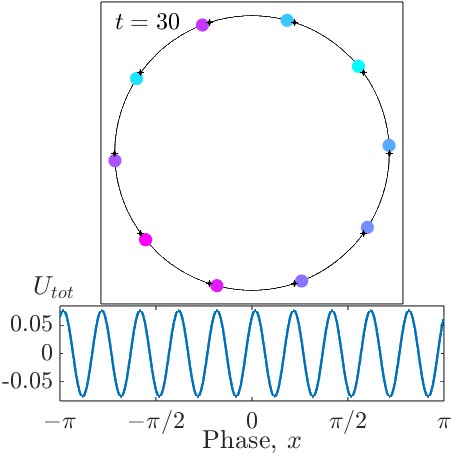}
	\caption{\textbf{Oscillator interpretation, slow convergence.} Ten Ricker oscillators with preferred distance $2\pi/10$. The left figure shows the system at time $t=5$, central figure at $t=25$, and right figure at $t=30$; the population swiftly self-arranges to become near-evenly but too-compactly arranged, then slowly separates, until suddenly ``snapping'' to perfectly even spacing. Black dots around circle indicate preferred distance; we see the particles eventually space themselves at the same intervals. Color indicates particle index, to distinguish and keep track of them over time.  }
	\label{fig:oscillators_slow_uniform}
\end{figure*}

Using the simpler, shortest-distance interpretation, we performed simulations and demonstrate the results in Fig.~\ref{fig:oscillators_slow_uniform}. We found that for small $n$, the particles did settle uniformly at intervals of $s$. Sometimes the system took a long time to find this state, as it evolves more slowly when evenly spaced---even when the population is more compact than necessary.

% \begin{figure}[ht!] 
% \centering
% 	%\includegraphics[width=.45\columnwidth]{512x2_overlay.jpg}
% 	\includegraphics[width=.45\columnwidth]{M6_T2.jpg}
% 	\caption{\textbf{Oscillator interpretation.} Six Ricker oscillators with preferred distance $2\pi/6$, at equilibrium when evenly spaced around the circle. Black dots around circle indicate preferred distance. Color indicates index, solely to distinguish particles. } 
% 	\label{fig:oscillators}
% \end{figure}

\begin{figure*}[ht!] 
\centering
	\includegraphics[width=.45\textwidth]{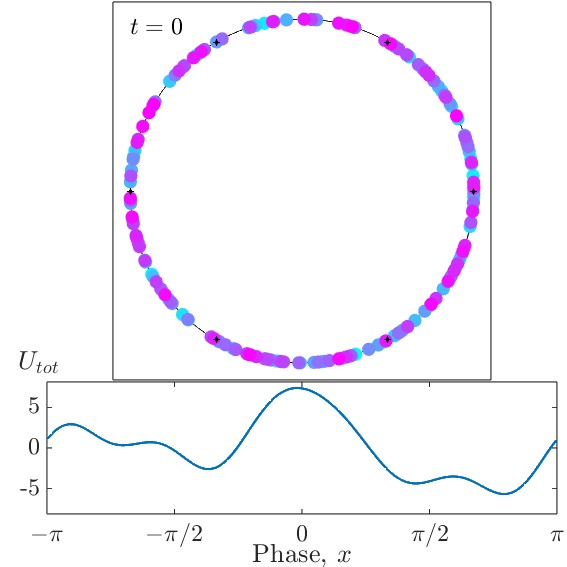}
    \includegraphics[width=.45\textwidth]{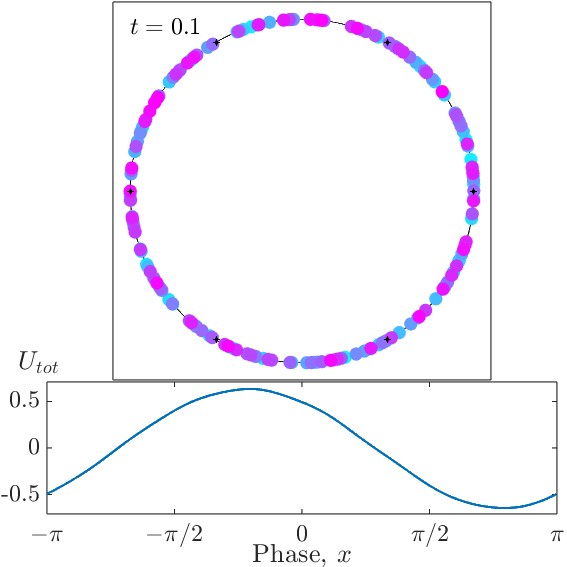}\\
    \includegraphics[width=.45\textwidth]{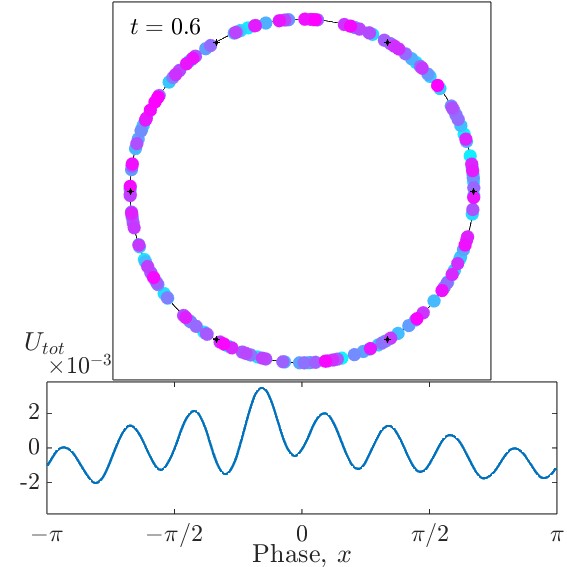}
    \includegraphics[width=.45\textwidth]{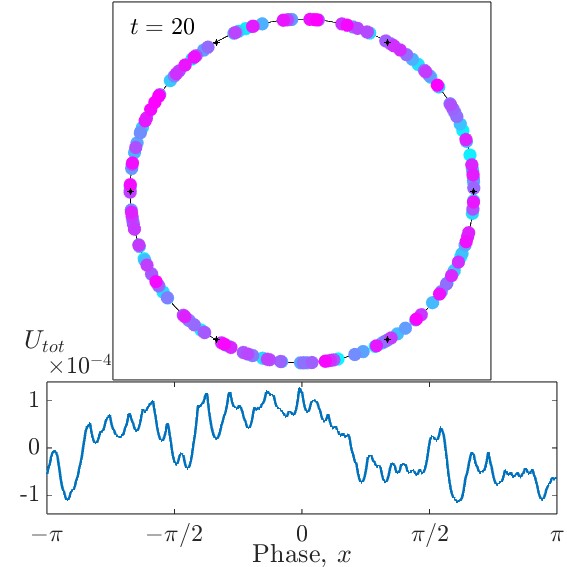}
	\caption{\textbf{Ricker oscillator system.} Two hundred Ricker oscillators with preferred distance $2\pi/6$. \textbf{Top left:} Random starting state, and its resulting cumulative potential. \textbf{Top right:} At $T=0.1$, the particles have very quickly arranged themselves into a single low-frequency cumulative-potential wave. \textbf{Bottom left:} At $T=0.6$, the low-frequency wave has damped, leaving a mid-frequency wave (with period $2\pi/8$, higher frequency than the Ricker wavelet's preferred distance) of much lower amplitude (two orders of magnitude smaller), with only minute positional adjustments. \textbf{Bottom right:} At $T=20$, the global potential has damped another order of magnitude, to $1 \times 10^{-4}$, leaving only the high-frequency spikes of individual Ricker wavelets (which have a ``sharp'' nondifferentiable corner as they wrap around $\pm \pi$). This ``glassy'' and distinctly nonuniform state appears to be stable, though it might only be extremely slow to evolve.} 
	\label{fig:oscillators_large_n}
\end{figure*}

However, at large $n$ (such that $n$ times the preferred distance was much larger than $2\pi)$, the system appeared to exhibit ``frozen disorder,'' or a ``glassy'' state where particles neither clump nor uniformly distribute (see Fig.~\ref{fig:oscillators_large_n}). The best lens for understanding this process appears to be the \textit{cumulative} potential; as the population evolves, it appears to self-organize almost instantly into a single low frequency wave (created by many individual Ricker potentials) which then damps quickly to reveal middle frequencies at smaller amplitude. As the magnitude of this cumulative potential wave shrinks beneath the scale of a single wavelet, the inherent higher frequencies of individual particles emerge again (see Fig.~\ref{fig:oscillators_large_n}, bottom right). 

This apparent phenomenon of self-organization in service of the cumulative potential's frequency-damping is only a numerical observation thus far, and merits future analytical exploration. It is unclear if this disordered state is truly stable or merely quasi-stable, and how the parameter-space transition from even-spacing to disordered ``equilibrium'' occurs as the domain becomes overpopulated relative to the preferred distance. We believe this is a ripe area for future exploration. 

\end{document}